\documentclass{elsart}
\usepackage{graphicx}

\newcommand{\req}[1]{(\ref{#1})}
\newcommand{\be}{\begin{equation}}
\newcommand{\ee}{\end{equation}}
\newcommand{\bea}{\begin{eqnarray}}
\newcommand{\eea}{\end{eqnarray}}

\newcommand{\avg}[1]{\langle{#1}\rangle}

\newcommand{\BE}{\begin{eqnarray}}
\newcommand{\EE}{\end{eqnarray}}
\newcommand{\BEn}{\begin{eqnarray*}}
\newcommand{\EEn}{\end{eqnarray*}}
\newcommand{\barr}{\begin{array}}
\newcommand{\earr}{\end{array}}

\newcommand{\bit}{\begin{itemize}}      
\newcommand{\eit}{\end{itemize}}
\newcommand{\bc}{\begin{center}}
\newcommand{\ec}{\end{center}}
\newcommand{\ben}{\begin{enumerate}}    
\newcommand{\een}{\end{enumerate}}

\begin{document}

\begin{frontmatter}
\title{Minority mechanisms in models of agents learning collectively a
  resource level}
\author{Damien Challet}
\address{Nomura Centre for Quantitative Finance, Mathematical Institute, Oxford University, 24--29 St Giles', Oxford OX1 3LB, United Kingdom}
\begin{abstract}
Starting from the Minority Game and building more and more sophisticated models of adaptive agents, we show that minority mechanisms underly any model where agents learn collectively a resource level that can be either obvious and constant in time, obvious and time-varying, or hidden.
\end{abstract}
\end{frontmatter}


Six years after its introduction, the Minority Game \cite{CZ97} is understood as a model of competition for limited resource. Many extensions of the original model have been introduced \cite{web}. remarkably, most of them contain a phenomenology which, while being partly idiosyncratic, has close similarities with that of the original MG. This is also true of models which relax the symmetry between the two choices, such as the El Farol bar problem \cite{Arthur,JohnsonAsym,CM03} and an asset pricing model \cite{Asset}. This also suggests that the original MG belongs to a broader class of models. For instance, the minority game (MG) was introduced as a simplified, binary version of the El Farol Bar problem (EFBP) \cite{Arthur}. Extending methods and results for the MG allowed recently the understanding of the EFBP \cite{CM03}. This raises two questions: what is these class of models, and how far can the methods used for the MG be extended to more general models.

Let us start from the MG, which is defined as follows: there are two choices, labelled by $-1$
and $+1$; at each time step, agent $i$ ($i=1,\cdots,N$) takes the choice $a_i(t)$. Then he receives the payoff 
\be
-a_i(t)A(t)
\ee
 where
$A=\sum_{i=1}^N a_i(t)$ is the aggregate outcome. He is therefore rewarded if he happened to be in the minority. The structure of a minority payoff is clearly made up of a $-$ sign and a product of the individual action and the aggregate action. When the agent possess strategies and associated cumulated scores, they are able to learn and adapt to their environment (see the original paper \cite{CZ97}. Many interesting issues and very rich behaviour were studied extensively in the past, and we shall refer the interested reader to previous work \cite{web}.

In the MG, the number of winners is at most $N/2$, which fixes the resource level $L$ at $N/2$. There is {\em a priori} no reason why $L$ should be fixed to $N/2$, except symmetry and simplicity. But a game with $L\ne N/2$ cannot reasonably be called a minority game: it is rather closer to an El Farol Bar problem. On the other hand, it is rather obvious that a common and essential mechanism is at work whatever $L$, which is called a minority mechanism in the following. Recognizing this allows the extension of the extensive literature on the MG to such games. 

The El Farol bar makes $N=100$ agents compete for $L=60$ seats. The economics literature has tried for many years now to try to understand why the average number of adaptive agents in the bar fluctuates around the resource level \cite{EF1,EF2}. In this problem, the (linear) payoff of agent $i$ is 
\be\label{EFpayoff}
-a_i(t)[A(t)-K] 
\ee 
where $K=|N-2L|$. The average attendance is equal to the resource level if
$\avg{A}=K$. In the MG, $K=0$, $\avg{A}=0$ by construction. It turns
out that $\avg{A}=K$ provided that the strategy space is
`consistent'\footnote{More precisely, a strategy space is consistent
  if $N \avg{a}_a =K$ where $\avg{a}_a$ denotes the average decision over the
  strategy space. See Ref. \cite{CM03} for more details.},
which is the case in Arthur's paper. Therefore, EFB is rigorously
equivalent to MG. As a consequence, all the MG literature's analysis of
fluctuations and predictability in the MG applies directly to the EFBP. Reversely, this explains why the phenomenology of the asset model \cite{Asset} is remarkably similar to that of the MG: in this model, $N$ agents try to discover the true value of an asset by buying and selling it, having only incomplete information. On
the other hand, this result also implies that that a game with $K\ne
0$ (or $L\ne N/2$) where the agents have binary strategies drawn in a non-biased way is not equivalent to a EFBP; accordingly, these models tend actually to study the by-products of strategy inconsistency whose effect is particularly dramatic with binary strategies. 

\begin{figure}
\centerline{\includegraphics[width=0.8\textwidth]{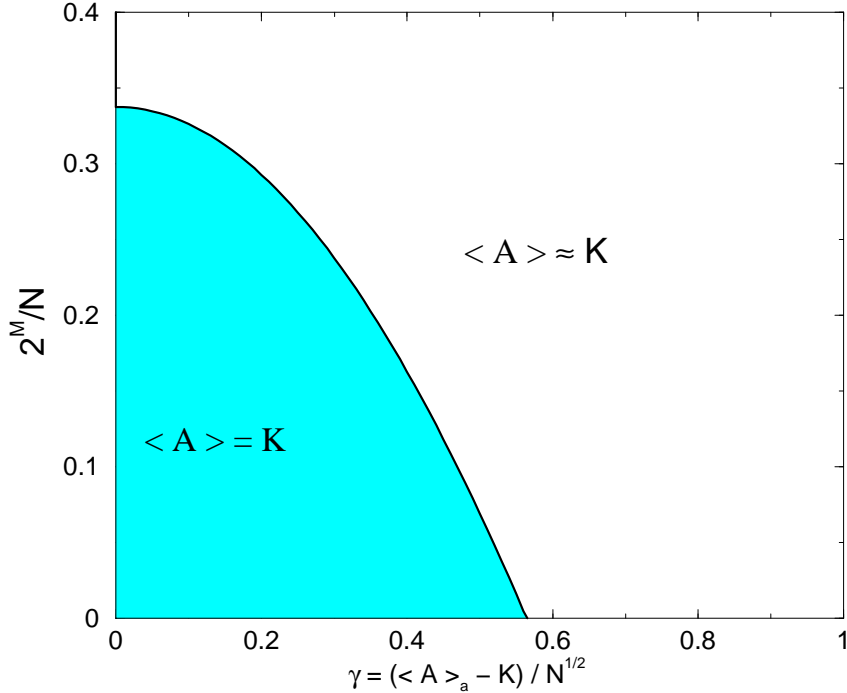}}
\caption{Phase diagram of the El Farol Bar problem and of Minority games with biased strategies. $\gamma$ measures the bias between the average element of the strategy space and the resource level.}
\label{phasediag}
\end{figure}

This makes it possible to compute the phase diagram of the EFBP when the agents based their decisions on previous $M$ (binary) attendances, as in the original EFPB: for perfect
strategy consistency (the $\gamma=0$ vertical line in Fig
\ref{phasediag}), $\avg{A}=K$. This plot illustrate the subtlety of the convergence of the attendance to the resource level: when $N$ is small compared with $P$ (upper part of the plot), a infinitesimal strategy inconsistency is enough to take $\avg{A}$ away from $K$. On the other hand, then $N$ is large enough, there is a region such that $\avg{A}=K$. It is however bounded: if $\gamma>1/\sqrt{\pi}$, no matter how many agents are in the game, the attendance never converges to the resource level.

The resource level $L$ can also vary in time. Instead of designing an {\em ad hoc} mechanism that changes $L$ according to some rule, be it deterministic or stochastic, but arbitrary nonetheless, let us consider commodity markets, or equivalently global ultimatum games \cite{COZ03} that perfectly show how minority mechanisms emerge when a group of players learn collectively a resource level. In a standard ultimatum game \cite{UG}, some generous but perverse donor proposes 1\$ to player 1, to be shared with player 2, provided that player 2 accepts the share offered by player 1. Rationality dictates that player 2 should accept any proposal, because otherwise he does not receive anything at all. But experiences show that a human player is not likely to propose, nor to accept an arbitrarily low proposal of player $i$, as they are not perceived as fair \cite{Roth}. Mathematically, player 1 proposes $a$ and player 2 expects at least $b$. Then player 1's payoff is 
\be
(1-a)\Theta(a-b)
\ee
and the one of player 2 is 
\be
a\Theta(a-b)
\ee
where $\Theta(x)=1$ if $x>0$ and $0$ otherwise. Suppose now that there is a group of $N_A$ people who have to share $N_{A}$\$ and a group of $N_{B}$ people who each expect to have a given amount of money. Mathematically, player $i$ of group $A$ offers $a_i$, while player $j$ of group $B$ expects at least $b_j$. The offers and expectations are grouped into
\be
A=\sum_{i=1}^{N_{A}}a_i \textrm{~~and~~}B=\sum_{j=1}^{N_{B}}b_j.
\ee
Then a transaction takes place if $A>B$. In that case, player $i$ of group $A$ receives $a_i(N_A-A)/N_A$, and player $j$ of group $B$ receives $b_iB/N_B$. Consider the very likely situation where $A\ne B$. Two types of player can be distinguished: those who want to see $A$ and $B$ converge, and those who try to make them diverge. The nice players of group $A$ have therefore a payoff that rewards small $a$ if $A>B$ and large $a$ if $A<B$; conversely for group-$B$ players, for instance, player $i$ of group $A$ rewards decision $a_i$ according to
\be
(1-a_i)\Theta(A-B)+a_i(1-\Theta(A-B))=a_i[1-2\Theta(A-B)]+\Theta(A-B)
\ee
Linearising this payoff gives $-a_i(A-B)+(A-B)$. Now, if player $i$ has $S>1$ strategies $a_{i,1},\cdots,a_{i,S}$, he will use them according to the difference between their cumulated payoffs. We can therefore drop the last term $(A-B)$; we conclude that strategies are rewarded according to
\be
-a_{i,s}(A-B)~~~~s=1,\cdots,S
\ee
Similarly, player $j$ of group $B$ rewards his strategies with $-b_{i,s}(B-A)$.
This is nothing else than a minority mechanism with resource level $B$ (see Eq \req{EFpayoff}). Therefore, $B$ is the resource level of the $A$ group, and $A$ is the resource level of the $B$ group: commodity markets are two coupled minority games, whose resource levels vary in time. The fact that each group plays internally a minority game means that the players of a given group compete with each other. For instance, in group $A$, each agent has incentive to lower his offer, hoping that his fellows will be generous enough to ensure that $A>B$. But if his fellows offer too little, he has better to offer more. Exact results of EFPB are readily extended to models with this kind of payoff.

A final example of well hidden minority mechanism is provided by credit risk models where $N$ banks have lent money to a given company \cite{CCh03}. The dynamics of the wealth $W$ of the latter is assumed to evolve according to
\be
\frac{W(t+1)}{W(t)}=1+\Delta(t)-\frac{B(t)}{W(t)}
\ee
where $\Delta(t)$ is the difference of relative wealth due to the activity of the company\footnote{It is often assumed to be a Gaussian variable of average $\mu$ and variance $\sigma$ for the sake of simplicity.} and $B(t)$ is the cost of the company's debt. More precisely, if bank $i$ lent $c_i$ and asks an interest rate of $r_i(t)$, $B(t)=\sum_{i=1}^Nc_ir_i(t)$. This type of equation is used in structure models of credit where the company is assumed to default if $W$ falls below a given threshold \cite{MertonStruct,BlackCox}. Although the major problem with this approach is obviously that $W$ is not observable, we shall let aside such considerations, as our point here is to show the existence of a minority mechanism. Assume that each bank has full liberty to fix its own interest rate. How should it react 
to bad news ($W<0$)? A decrease in the wealth means that the company is more likely to default in the future, hence the banks usually tend to increase their interest rate. But doing so, they also increase the risk of default. Reversely, if a bank decreases its interest rate when $W$ decreases and increases it when $W$ increases, it rewards interest rate $r_i$ according to
\be
r_i\frac{W(t+1)-W(t)}{W(t)}=b_i\Delta(t)-\frac{r_i\sum_{j=1}^Nr_j(t)c_j}{W(t)}
\ee
The term $-r_i\sum_{j}r_ic_j$ reveals a minority mechanism; conversely, usual banks play a majority game. The resource level is implicitly fixed by the condition of zero average wealth change $\avg{\Delta} W=\avg{\sum_i r_ic_i}$. This means that as long as the players can learn this resource level, $W$ will oscillate around an average value. As a consequence, the company is not likely to go bankrupt, but cannot grow either: minority players are good for the company when it does badly, but bad when it does relatively well; when the company does very well, that is, when $\avg{\Delta}W\gg \avg{\sum_i r_ic_i}$, the minority-player banks ask for the maximum rate, but credit repayment has a negligible importance, hence the company can grow. The competition between minority players is of the same kind as in commodity markets: for them, it is tempting to ask for a larger interest rate and hope that other banks ask for a small rate; on the other hand, if many banks ask for too high a rate, some others have to lower their rate else the risk of default is too high. 

In conclusion, reviewing more and more complex situations with adaptive agents leads to the conclusion that minority mechanisms are found when some agents learn collectively a resource level, hidden or obvious, constant in time or not, going beyond the usual belief that minority games capture only competition for a scarce resource. Adaptive agents act so as to cancel the average payoff, which defines the resource level implicitly, and leads to a mean-reverting process. Although exact analytical methods used for the standard minority game can be extended to many simple extensions of the MG, how far this can be done is an open question.




\begin{thebibliography}{99}
\bibitem{CZ97} D. Challet and Y.-C. Zhang, Physica A {\bf 246}, 407
(1997) preprint adap-org/9708006
\bibitem{web} D. Challet, Minority Game web site, {\tt www.unifr.ch/econophysics/minority}
\bibitem{Arthur} Arthur W. B., 
Am. Econ. Assoc. Papers and Proc {\bf 84}, 406, 1994.
\bibitem{JohnsonAsym} N.F. Johnson {\em et al.}, Physica A {\bf 269}, 493 1999
\bibitem{CM03} D. Challet, G. Ottino and M. Marsili, Physica A {\bf 332}, 469 (2004), preprint cond-mat/0306445
\bibitem{Asset} J. Berg {\em et al.}, Quant. Fin {\bf 1}, 203--21, preprint cond-mat/0101351
\bibitem{EF1} J. L. Casti, Complexity {\bf 1}, 7 (1996)
\bibitem{EF2} D. B. Fogel, K. Chellapilla, P. J. Angeline, IEEE
  Trans. Ev. Comp. {\bf 3}, 142 (1999)
\bibitem{COZ03} D. Challet, G. Ottino and Yi-Cheng Zhang, in preparation
\bibitem{UG} W. G\"uth, R. Schmittberger, B, Schwarze, 
 J. Econ. Behav. Organ. {\bf 3}, 376--388 (1982).
\bibitem{Roth} A. E. Roth, Bargaining experiments. in {\em Handbook of experimental economics}, ed. J. Kagel and A. E. Roth, 171--202, Princeton University Press, 1995
\bibitem{CCh03} D. Challet and M. Chromik, in preparation
\bibitem{MertonStruct} R. Merton, C. Robert, J. Fin. {\bf 29}, pp. 449-470 (1974)
\bibitem{BlackCox} F. Black, J. C Cox, J. Fin. {\bf 31}, 351--367 (1976)
\end{thebibliography}
\end{document}